\documentclass[a4paper]{revtex4}
\usepackage{graphicx}
\usepackage{fancyhdr}
\usepackage{amsmath}
\newcommand{\order}[1]{{\cal O}(#1)} 

\newcommand{\TeV}{\mbox{TeV}}
\newcommand{\GeV}{\mbox{GeV}}
\newcommand{\MeV}{\mbox{MeV}}

\usepackage{color}

\begin{document}

\title{Phenomenological constraints on light mixed sneutrino WIMP scenarios
\footnote{This report is based on collaboration with Mitsuru Kakizaki, Eun-Kyung Park, and Jae-hyeon Park \cite{Kakizaki:2015nua}.
}}

%

  \author{Akiteru Santa} 
  \affiliation{
  Department of Physics,
  University of Toyama, Toyama 930-8555, Japan
  }

\begin{abstract}
Scenarios where left- and right-handed sneutrinos mix and 
the lightest mixed one act as a thermal dark matter candidate 
can solve the dark matter, neutrino mass, and hierarchy problems simultaneously.
We focus on the dark matter mass region of the order of $1$ GeV, where dark matter direct detections are insensitive. 
We calculate the decay rate of a false vacuum induced by a large 
sneutrino trilinear coupling as well as other observables constrained by experiments.
As a result, we show that there is an allowed region where the mass of the lightest neutralino is 
around $1$ GeV.
The allowed region can be tested by the search for the Higgs boson invisible 
decay at future colliders.
\end{abstract}

\maketitle

\thispagestyle{fancy}


\section{Introduction}
In July 2012, the ATLAS and CMS collaborators reported the discovery of a new particle 
with a mass of $125$ GeV\cite{LHC}.
Data revealed that the phenomenological profile of the new particle resembles that of the Higgs 
boson of the Standard Model (SM).
The SM has been established as an effective theory at low energy scales below $\mathcal{O}(100)\ \mathrm{GeV}$.
However, there are some unsolved problems.
In the SM, neutrinos are massless particles although neutrino oscillation phenomena were discovered\cite{SK}.
The SM does not contain any candidate which can explain the dark matter relic abundance\cite{WMAP,Ade:2013zuv}.
In the theoretical view point, the SM has the problem that quantum corrections to the squared Higgs boson mass diverge quadratically.

A weakly interacting massive particle (WIMP) is a good dark matter candidate.
The relic density and the annihilation cross section of the WIMP are related to each other if the WIMP was in thermal equilibrium in the early universe.
In order to reproduce the observed relic density, the annihilation cross section of $\order{\mathrm{pb}}$ is required.
The cross section implies the new physics energy scale of $\order{1\TeV}$ according to the dimensional analysis.
These parameter regions are explored by dark matter direct and indirect detections and the Large Hadron Collider (LHC).

In this paper, we focus on the model extended by introducing right-handed neutrinos and supersymmetry (SUSY).
The right-handed neutrinos produce Dirac neutrino masses.
The extended model with SUSY particles can avoid the hierarchy problem because of the cancelations between the quantum corrections to the squared Higgs boson mass from the SM particles and those from their superpartners.
The superpartners of the right-handed neutrinos, right-handed sneutrinos can mix with left-handed ones.
 If the lighter mixed sneutrino is the lightest SUSY particle (LSP), the sneutrino acts as a WIMP candidate\cite{ArkaniHamed:2000bq,Belanger:2010cd,Dumont:2012ee}.
Earlier works analyze parameter regions with the gaugino mass universality and find that there are allowed regions where the mixed sneutrino dark matter mass is heavier than half of the Higgs boson one.
However, the lighter mass regions are excluded by the limits of the relic density and the branching ratio of the Higgs boson invisible decay.

We explore the GeV-mass mixed sneutrino dark matter scenarios without the gaugino mass universality.
In such a region, 
dark matter direct detections are insensitive because of their energy thresholds, and 
a large sneutrino trilinear coupling triggers a deeper vacuum than the SM-like one.
We calculate the decay rate of the false vacuum neglected in earlier works and impose the vacuum stability bound on parameter space.
We show that there is a region consistent with all phenomenological constraints, and the allowed region can be examined by the search for the Higgs boson invisible decay at future colliders.

\section{Model}
The mixed sneutrino model contains right-handed neutrinos $\nu_{Ri}$ and sneutrinos $\widetilde{\nu}_{Ri}$ in addition to the usual particles of the Minimal Supersymmetric Standard Model (MSSM).
Here, the index $i=1,\ 2,\ 3$ denotes the generation.
Neutrino Yukawa interaction, sneutrino soft mass, and sneutrino trilinear coupling terms are introduced to the MSSM Lagrangian. 
The soft masses and the trilinear couplings among the right-handed sneutrino, the left-handed slepton doublet $\widetilde{\ell}_i$, and the Higgs doublet with hypercharge $Y=+1/2$, $h_u$ are written as
\begin{eqnarray}
  \Delta {\cal L}_{\rm soft} = m^2_{\widetilde N_i}  |\widetilde \nu_{Ri} |^2 +  
        A_{\tilde\nu_i} \widetilde \ell_i \widetilde \nu_{Ri}^* h_u + {\rm h.c.} \,,
\end{eqnarray}
where $m_{\widetilde N_i}$ denote the soft mass parameters of the right-handed sneutrinos,
and $A_{\tilde\nu_i}$ are the sneutrino trilinear coupling constants.
After the Higgs bosons develop vacuum expectation values,
the couplings contribute to non-diagonal components of the sneutrino mass matrix.
The mass matrix for one generation is given by
\begin{eqnarray}
{\cal M}^2_{\tilde\nu} =
 \left( 
\begin{array}{cc}
  {m}^2_{\widetilde{L}} +\frac{1}{2} m^2_Z \cos 2\beta  &  \frac{1}{\sqrt{2}} A_{\tilde\nu}\, v \sin\beta\\
  \frac{1}{\sqrt{2}}    A_{\tilde\nu}\, v \sin\beta&  {m}^2_{\widetilde{N}}
 \end{array}\right) \, ,
\label{eq:sneutrino_tree}
\end{eqnarray}
where $m_{\widetilde{L}}$ means the soft mass parameter of the left-handed sleptons.
The sum of the squared vacuum expectation values (the ratio of the vacuum expectation values) is written by $v^2 = v_1^2 + v_2^2$ ($\tan\beta = v_2/v_1$), where
$v_1\ (v_2)$ is the vacuum expectation value of the Higgs doublet with $Y=-1/2\ (Y=+1/2)$.
Therefore, the left- and right-handed sneutrinos mix and one obtains mass eigenstates,
\begin{eqnarray}
 \tilde\nu_1 = \cos\theta_{\tilde\nu} \, \tilde\nu_R - \sin\theta_{\tilde\nu}\,  \tilde\nu_L \,, \quad
 \tilde\nu_2 = \sin\theta_{\tilde\nu} \, \tilde\nu_R + \cos\theta_{\tilde\nu}\,  \tilde\nu_L .
\end{eqnarray}

In this paper, we consider the third generation mixed sneutrino WIMP scenarios assuming that the first and second sneutrinos are heavier than any experimental limit.
\section{Constraints}
We discuss phenomenological constraints on the GeV-mass sneutrino WIMP scenarios.
We use the experimental results listed in TABLE \ref{tab:expresulte} in order to analyze the parameter space.
\begin{table}
\begin{center}
\caption{Observables and experimental constraints.}
\begin{tabular}{|c||c|c|}
	\hline
	Observable & Experimental result\\ \hline \hline
	$\Omega h^2$ & $0.1196 \pm 0.0062\ (95\%\ \mathrm{CL})$ \cite{Ade:2013zuv}\\  \hline
		$\sigma_{\rm N}^{\rm SI}$ & $(m_{\rm DM},\ \sigma_{\rm N}^{\rm SI})$ constraints \\
& from LUX \cite{Akerib:2013tjd} and SuperCDMS \cite{Agnese:2014aze} \\ \hline
	$\sigma_{\rm ann} v$  & $(m_{\rm DM},\ \sigma_{\rm ann}v)$ constraint \\
	& from FermiLAT \cite{Ackermann:2013yva} \\ \hline
	$\Delta \Gamma (Z \rightarrow \mathrm{inv.} )$ & $< 2.0\ \MeV \ (95\%\ \mathrm{CL})$
		 \cite{ALEPH:2005ab} \\ \hline
	$\mathrm{Br}(h \rightarrow \mathrm{inv.} )$ & $< 0.29 \ (95\%\ \mathrm{CL})$ \cite{ATLAS-CONF-2015-004} \\ \hline
	$m_{\tilde{\tau}_R}$ & $> 90.6\ \mathrm{GeV} \ (95\%\ \mathrm{CL})$ \cite{Aad:2014yka} \\ \hline
	$ m_{\widetilde{\chi}^{\pm}_1}$ & $> 420\ \mathrm{GeV} \ (95\%\ \mathrm{CL})$ \cite{Aad:2014yka} \\ \hline
	$ m_{\tilde{g}}$ & $> 1.4 \ \mathrm{TeV} \ (95\%\ \mathrm{CL})$
		\cite{Aad:2014lra, Chatrchyan:2014lfa} \\ \hline
\end{tabular}
\label{tab:expresulte}
	\end{center}
\end{table}
%
The GeV-mass sneutrino dark matter tends to annihilate via neutralinos.
Then, the annihilation cross section depends on the neutralino masses as well as the sneutrino mixing angle.
For $M_{\widetilde{B}}\ll M_{\widetilde{W}}$,
the relic abundance is given approximately by 
\begin{equation}
	\Omega h^2 \sim 0.1 \times 
	\left ( \frac{\sin \theta_{\tilde{\nu}}}{0.1} \right)^{-4}
	\left( \frac{m_{\tilde{\chi}^0_1}}{1\ \mathrm{GeV}} \right)^2\, .
\label{eq:rough omega}
\end{equation}
If the lightest neutralino mass is around $1$ GeV and the mixing angle is around $0.1$,
one obtains the correct relic abundance.

Coupling constants of dark matter candidates are constrained through direct and indirect detections\cite{Akerib:2013tjd,Agnese:2014aze,Ackermann:2013yva}.
However, direct detections are insensitive to GeV-mass region because of their energy thresholds.
Indirect detections which observe charged particles and photon from dark matter annihilations cannot impose limits on the GeV-mass sneutrino dark matter.
 
Let us turn to constraints from collider experiments.
The invisible decays of the $Z$ and Higgs bosons are explored by LEP2 and the LHC experiments, respectively\cite{ALEPH:2005ab, ATLAS-CONF-2015-004}.
In our model, the decay rates are proportional to $\sin^4\theta_{\widetilde{\nu}}$.
Then, the experimental limits impose upper limits on the sneutrino mixing angle.
Productions of electroweak superparticles at the LHC are associated with signals with two or three leptons.
In our model, the lightest chargino (next-to-lightest neutralino) decay to one tau (two taus) with missing energy.
Therefore, we impose the constraint from the search for two or three taus\cite{Aad:2014yka}.
LEP2 and the LHC experiments search for a mono-photon event with missing energy also\cite{Achard:2003tx,Aad:2014tda}.
Such constraints are not serious since the cross section is enough small in our model\cite{Belanger:2010cd}.

In the MSSM, a large trilinear soft SUSY breaking term triggers a deeper vacuum than the SM-like one\cite{ccb}.
In our model, the sneutrino trilinear coupling is large.
Through tracing the scalar potential along the $D$-flat direction, $|h_u^0| = |\tilde{\nu}_L| = |\tilde{\nu}_R| = a$,
$\theta_{\tilde{\nu}} \gtrsim 2\times 10^{-12}$
suggests a lepton-number breaking global minimum.
We calculate the decay rate of the false vacuum and check whether the lifetime of the universe is enough long.
The vacuum meta-stability bound impose the upper limit, $\theta_{\tilde{\nu}}\leq 0.52$ for $m_{\tilde{\nu}_1} = 0.1\ \GeV$.
The upper limit is relaxed in the larger mass region since the trilinear coupling constant is proportional to $m_{\tilde{\nu}_2}^2-m_{\tilde{\nu}_1}^2$.

\section{Analysis}
\begin{figure}[]
\includegraphics{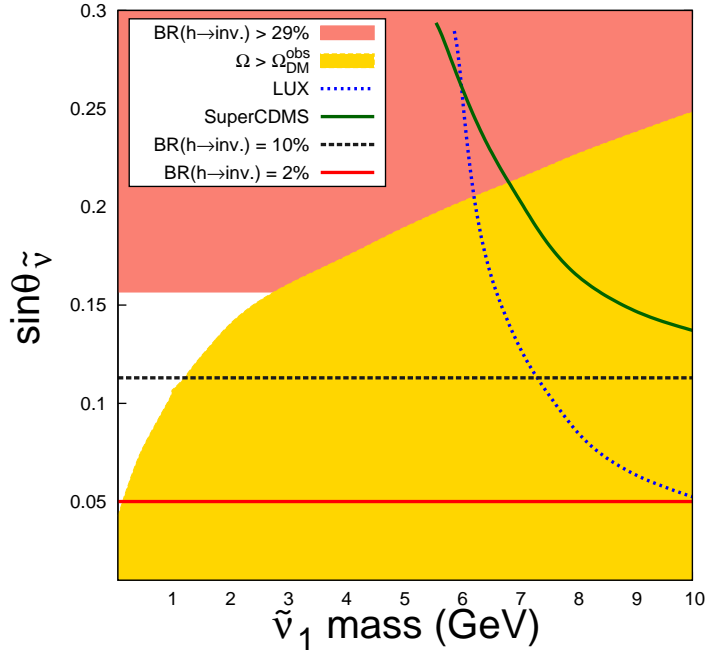}
\caption{\footnotesize The results of our parameter scan for light
  mixed sneutrino WIMP scenarios in the $(m_{{\tilde \nu}_1},
  \sin \theta_{\tilde{\nu}})$ plane.  The yellow (light-gray) and pink
  (dark-gray) regions are excluded by the constraints of the relic
  abundance \cite{Ade:2013zuv} and the Higgs boson invisible decay
  \cite{ATLAS-CONF-2015-004}, respectively.  We also show the upper
  limits of the spin-independent elastic WIMP-nucleon cross section by
  the LUX (blue dotted line) \cite{Akerib:2013tjd} and the SuperCDMS
  (dark-green line) \cite{Agnese:2014aze}.  The black dashed (red
  solid) line denotes the Higgs boson invisible decay branching
  fraction of $10\%$ ($2\%$).  }
\label{fig:general1g}
\end{figure}

We perform a scan for the parameter region as listed in TABLE \ref{tab:inputparameter}, 
and show the phenomenological constraints in the $(m_{\tilde{\nu}_1},\ \sin\theta_{\tilde{\nu}})$ plane in FIG. \ref{fig:general1g} \cite{update}.
\begin{table}[]
	\begin{center}
	\caption{Parameters and reference values/scan bounds. 
	}
	\label{tab:inputparameter}
	\begin{tabular}{|c||c|c|}
	\hline
	Parameter & Reference value/Scan bound\\ \hline \hline
	$ \mu $ & $500\ \mathrm{GeV} $ \\ \hline
	$ \tan \beta$ & $10 $ \\ \hline
	$m_{\tilde{\nu}_2}$ & $125\ \mathrm{GeV} $ \\ \hline
	$m_{\tilde{\tau}_R}$ & $120\ \mathrm{GeV}$ \\ \hline
	$M_{\widetilde{W}}$ & $500\ \mathrm{GeV}$ \\ \hline
	$m_{\tilde{\nu}_1}$ & $[0.1\ \mathrm{GeV},\ 10\ \mathrm{GeV}] $ \\ \hline
	$\sin \theta_{\tilde{\nu}}$& $[0.01,\ 0.3] $ \\ \hline
	$M_{\widetilde{B}}$ & $[0.1 \ \mathrm{GeV} ,\ 20\
	\mathrm{GeV} ] $ \\ \hline
	\end{tabular}
	\end{center}
\end{table}
The colored regions are ruled out by the constraints of the relic abundance and the Higgs boson invisible decay.
The white region is consistent with the vacuum stability bound as well as the experimental constraints.
The allowed region will be narrowed by searches for the Higgs boson invisible decay at future colliders.
The high-luminosity LHC with the center-of-mass energy of $\sqrt{s}=14\ \TeV$ and the luminosity of  $L=3000\ \mathrm{fb}^{-1}$can impose the upper limits of $\mathrm{Br}(h \rightarrow \mathrm{inv.}) <8.0\%\ (95\%\ \mathrm{CL})$ \cite{ATL-PHYS-PUB-2013-014} and 
$\mathrm{Br}(h \rightarrow \mathrm{inv.}) <6.4\%\ (95\%\ \mathrm{CL})$\cite{CMS:2013xfa} .
The International Linear Collider (ILC) can constrain the branching ratio up to $0.69\%\ (95\%\ \mathrm{CL})$\cite{Ishikawa}.
The ILC is capable of excluding mixed sneutrino dark matter scenarios for $0.1\ \mathrm{GeV} \leq m_{\tilde{\nu}_1} < 3\ \mathrm{GeV}$.

\section{Conclusion}
We have analyzed GeV-mass mixed sneutrino dark matter scenarios relaxing the gaugino mass universality and imposing the vacuum stability bound.
If the mass of the lightest neutralino is of the order of GeV, these scenarios are consistent with all the phenomenological constraints.
The allowed region can be probed through the searches for the Higgs boson invisible decay although dark matter direct detections are insensitive to the GeV-mass dark matter region.

\begin{acknowledgments}
The work presented here is done in collaboration with Mitsuru Kakizaki, Eun-Kyung Park, and Jae-hyeon Park to whom I express my deep thanks for fruitful collaborations.
\end{acknowledgments}
\bigskip 

\end{document}